\documentclass[journal]{IEEEtran}

\usepackage{cite}
\usepackage{amsmath,amssymb,amsfonts}
\usepackage{algorithmic}
\usepackage{graphicx}
\usepackage{subfigure}
\usepackage{textcomp}
\usepackage{makecell}
\usepackage{booktabs}
\usepackage{colortbl}

\begin{document}

\title{AI-Driven Radio Propagation Prediction in Automated Warehouses using Variational Autoencoders}

\author{
{
Rahul Gulia,
Amlan Ganguly,
Michael E. Kuhl,
Ehsan Rashedi,
and Clark Hochgraf
}
\thanks{\small R. Gulia and A. Ganguly are with the Department of Computer Engineering, Rochester Institute of Technology, Rochester, NY, USA (e-mail: rg9828@rit.edu).}
\thanks{\small M. E. Kuhl and E. Rashedi are with the Department of Industrial and Systems Engineering, Rochester Institute of Technology, Rochester, NY, USA.}
\thanks{\small C. Hochgraf is with Re:Build Manufacturing, USA.}
\thanks{\small This work is published in IEEE Transactions on Industrial Informatics. The final published version is available at DOI: 10.1109/TII.2026.3682204.}
}

\markboth{Journal of \LaTeX\ Class Files,~Vol.~14, No.~8, August~2015}%
{Shell \MakeLowercase{\textit{et al.}}: Bare Demo of IEEEtran.cls for IEEE Journals}

\maketitle

\begin{abstract}
The pervasive demand for data-intensive applications and the rapid integration of emerging technologies are driving an unprecedented transformation in wireless communication, particularly within Industry 4.0. Optimizing 5G and future networks for automated environments like smart warehouses requires advanced solutions for indoor radio propagation. To this end, this paper introduces WISVA (Wireless Infrastructure for Smart Warehouses using VAE), an AI-based framework utilizing a novel Variational Autoencoder (VAE) model (AI Contribution). The VAE's unique architecture learns complex electromagnetic (EM) wave interactions from meticulously crafted, physics-informed data tensors, enabling it to accurately model signal behavior impacted by diverse obstacles. This engineering application provides site specific signal-to-interference-plus-noise ratio (SINR) heatmaps with relative fine granularity for 5G wireless bands in automated Industry 4.0 settings. We demonstrate the remarkable robustness and adaptability of WISVA through its superior performance in spatial field reconstruction tasks, its validation, and, critically, its ability to extrapolate to entirely unseen warehouse layouts and configurations. Comparative analysis via reconstruction error heatmaps reveals WISVA's significantly higher accuracy against traditional autoencoders, establishing its potential as a critical enabler for efficient wireless infrastructure optimization in Industry 4.0.
\end{abstract}


%
\IEEEpeerreviewmaketitle

\section{Introduction }
\IEEEPARstart{T}{he} upcoming decade promises an unprecedented shift in wireless communication, driven by the mounting demand for data-rich and latency-sensitive applications central to Industry 4.0 \cite{Industry4_0}. This transformation is particularly vital for the efficient operation of automated warehouses, which require seamless, intelligent infrastructure to support assets like Autonomous Material Handling Agents (AMHAs) via vehicle-to-infrastructure (V2I) links \cite{V2I}. The sophisticated nature of these operations necessitates precise wireless connectivity planning and optimization across every area to ensure reliable, uninterrupted coverage and facilitate scalability.

The implementation of 5G and Millimeter-wave (mmWave) frequencies provides the necessary ultra-fast data transmission and low latency required by these systems. However, mmWave deployments face distinct indoor challenges: the high-frequency signals are highly susceptible to severe attenuation from common building materials and reflective obstacles, leading to complex propagation phenomena like multipath and shadowing \cite{mmWave_attenuation}. Currently, estimating wireless connectivity, specifically the Signal-to-Interference-plus-Noise Ratio (SINR), in these environments relies on either expensive channel measurements or time-consuming simulations using tools like Network Simulator-3 (ns-3) \cite{ns_3}. As shown in Table \ref{tab:Table_SimTime}, generating a relative fine granularity SINR heatmap for a complex warehouse can take nearly 38 hours on high-performance hardware, making traditional methods impractical for real-time network management or rapid optimization.

\begin{table}[htbp!]
\centering
\caption{ns-3 5G-LENA simulation time for SINR heatmaps of a 60$\times$60 m warehouse at various spatial granularities.}
\label{tab:Table_SimTime}
\renewcommand{\arraystretch}{1.2}
\setlength{\tabcolsep}{2pt} 
{\small
\begin{tabular}{|c|c|c|c|}
\hline
\textbf{Grid Granularity} & \textbf{Point Dist. (m)} & \textbf{Single AP} & \textbf{Full Warehouse} \\ 
\textbf{(pixels)} & & \textbf{(min:s)} & \textbf{(hr:min)} \\ 
\hline
50$\times$50 & 1.20 & 0:20 & 0:48 \\
100$\times$100 & 0.60 & 0:40 & 1:36 \\
200$\times$200 & 0.30 & 2:42 & 6:28 \\
450$\times$450 & 0.13 & 12:54 & 30:57 \\
500$\times$500 & 0.12 & 15:48 & 37:55 \\
\hline
\end{tabular}
}
\end{table}

The simulation results in Table \ref{tab:Table_SimTime} highlight a critical challenge for real-world implementation by showing the excessive time required for relative fine granularity SINR heatmaps \cite{Evaluation_60GHz}. To meet the demand for agile, scalable network solutions, we propose WISVA (Wireless Infrastructure for Smart Warehouses using VAE), an AI-based framework that leverages a Variational Autoencoder (VAE) \cite{variational_autoencoding}. Unlike conventional MLP or simple Autoencoder models that struggle with generalization and exhibit high error in complex indoor settings \cite{EM_DeepRay}, WISVA uses a physics-informed approach. Its multi-branch encoder models distinct physical tensors—specifically permittivity (material properties), Euclidean distance (geometry), and Access Point (AP) layout—to extract a disentangled, robust latent representation. This design ensures that the model accurately captures EM wave interactions and enables it to generalize across new and entirely unseen warehouse layouts, making it an effective tool for closed-loop network optimization.

In conclusion, this generative AI framework represents a significant advancement in indoor wireless network planning for high-frequency bands. It simplifies the design and optimization process, enhances prediction accuracy, and enables the feasible deployment of high-performance, resilient wireless networks in automated warehouses.

The contributions of this paper are summarized as follows: 
\begin{enumerate} 
    \item Introduction of WISVA for Complex Industry 4.0 Environments: We present WISVA, a VAE-based framework specifically designed to accurately model indoor radio propagation within intricate smart warehouse settings. 

    \item Integration of Physics-informed Tensors for Enhanced Modeling: WISVA leverages physics-informed tensors (permittivity, geometry, AP layout) to capture EM wave behaviors, enabling a higher degree of generalization across multiple warehouse configurations.

    \item Comprehensive Evaluation Across Diverse Scenarios: We rigorously assess WISVA's capabilities across critical use cases, including SINR Heatmap Prediction, spatial field reconstructions, and superior Extrapolation to entirely novel layouts.

    \item Superior Performance Over Baseline Models: Our extensive experiments demonstrate that WISVA outperforms traditional Autoencoders (AE) and state-of-the-art models in terms of accuracy, robustness, and efficiency. 
\end{enumerate} 

The remainder of this paper is structured as follows. Section II discusses related work concerning both traditional and AI-driven radio propagation models. Section III outlines the theoretical background of the VAE architecture. Section IV details the generation of the high-fidelity ns-3 simulation dataset used for training. Section V presents the complete WISVA model architecture and the training methodology. Section VI reports on the rigorous experimental results. Finally, Section VII concludes the paper and discusses future work.

\section{Related Work}

Earlier research conducted an extensive analysis of AMHA localization within smart warehouses, focusing on wireless technologies such as Ultra WideBand (UWB), 5G New Radio (NR), and WiGig (60 GHz). Precise localization was identified as a critical enabler for Industry 4.0, supporting worker safety and efficient AMHA navigation. Consequently, previous studies proposed ML-based techniques leveraging radio maps to estimate AMHA locations within these environments \cite{AutomatedWarehouse, Evaluation_60GHz}.

Building upon this foundation, a preliminary study \cite{WarehouseVAE} introduced an initial VAE-based model for indoor radio propagation. This paper significantly expands those findings, providing a more comprehensive analysis and deeper comparisons. Furthermore, while earlier work \cite{UAVPathLoss} investigated non-deterministic statistical models like the U-LAAG for 2.4-2.5 GHz networks, such approaches underscore the necessity for advanced, data-driven techniques like WISVA to accurately model the complex dynamics of 5G environments.

A variety of related works have explored data-driven wireless communication, using neural networks to predict coverage probability \cite{relatedwork_1} or reviewing AI applications for deployment and localization in WSNs \cite{relatedwork_6}. Other ML approaches for SINR prediction include LSTM and Random Forest models for outdoor private campus networks considering mobility \cite{relatedwork_2}, Kriging-based techniques for mobile networks \cite{relatedwork_3}, and IndoorRSSINet for generating 2D RSSI maps \cite{relatedwork_5}. However, these methods generally lack the necessary fidelity for complex indoor mmWave environments. 

In parallel, recent studies have explored hybrid machine learning and optimization frameworks for complex engineered systems, particularly in energy and cyber-physical domains. These works combine advanced neural architectures with metaheuristic optimization, attention mechanisms, and spectral analysis to improve learning efficiency and robustness \cite{Liu1, Liu2, Liu3, Liu4, Liu5, Liu7}. While these efforts primarily address temporal forecasting and system-level control, they reflect a broader trend toward physics-aware neural models. We cite these efforts to contextualize WISVA within this evolving landscape, noting that our focus differs by learning dense, spatially distributed radio propagation fields governed by electromagnetic interactions in complex indoor environments.

More recent efforts employ advanced neural field modeling, such as $\text{RadioUNet}$ \cite{Comparison_RadioUNet} and $\text{NeRF}^{2}$ \cite{Comparison_NERF2}. However, $\text{RadioUNet}$'s reliance on implicit geometric features limits its utility where material properties dominate signal behavior, and $\text{NeRF}^{2}$'s dense 3D rendering is impractical for real-time industrial prediction. Similarly, GNN models \cite{relatedwork_4} do not integrate physics-informed information about interacting objects, and SDU-net \cite{EM_DeepRay} depends on simplified planar physics models. 

In contrast, the WISVA model explicitly addresses these gaps by leveraging physics-aware tensors—including permittivity, AP locations, and Euclidean distance—to accurately capture EM wave interactions. Furthermore, it provides uncertainty quantification through probabilistic SINR distributions via the VAE's latent space, which is uniquely suited for stochastic indoor environments. This latent-space modeling ensures high adaptability and superior generalization to entirely new layouts compared to models requiring extensive retraining. Finally, recent studies have explored ML for operations such as path loss prediction in mmWave bands using ANNs and RNN-LSTM models \cite{WISVA_1}, and D2D communication variations using GRU models \cite{WISVA_2}. WISVA integrates these advancements into a novel framework specifically tailored to the unique challenges of Industry 4.0.


\section{Theoretical Background}

A Variational Autoencoder (VAE) is a generative AI framework that utilizes deep learning for content generation, anomaly detection, and denoising \cite{VAE_1}. By merging standard autoencoding principles \cite{AE_6} with probabilistic modeling, VAEs acquire structured latent representations that enable the traversal from known to unknown feature domains during inference. Unlike deterministic models, VAEs learn the intrinsic probability distribution of input data, allowing them to sample and generate novel instances similar to the training set.

A primary strength of VAEs is their ability to learn disentangled latent representations \cite{VAE_2}, where individual dimensions correspond to specific semantic or physics-informed features—such as structural obstructions or AP proximity. This flexibility allows VAEs to reconstruct multi-dimensional datasets and impute missing values in complex environments with varying resolutions. 

The WISVA framework leverages these capabilities to compress detailed warehouse blueprints into manageable latent manifolds that highlight key propagation influences. By revealing underlying factors like wall materials and equipment placement, WISVA predicts signal quality in uncharted areas by generating high-fidelity data points where direct measurements are unavailable. This predictive capacity ensures reliable wireless coverage, facilitating the seamless operation and improved productivity of warehouse automation systems.


\section{Training Dataset Generation for WISVA} \label{Training_Dataset_Generation}

The WISVA model relies on a dataset of synthetic SINR heatmaps representing various indoor 5G environments, generated via the \text{ns-3} 5G Lena module \cite{5GLena}. Training data quality and quantity directly dictate the model's effectiveness. While the current WISVA implementation is optimized for the 60~GHz (V-band) regime due to its severe shadowing characteristics, the framework is architecturally frequency-agnostic; adaptation to other mmWave bands, such as 28~GHz, simply requires updating the material attenuation parameters within the permittivity tensors.

A systematic approach captures a vast manifold of electromagnetic (EM) interactions by translating the AP across a high-density $5\text{m} \times 5\text{m}$ grid within five architectural templates. At 60~GHz, each AP shift creates a unique diffraction and shadowing profile relative to the 20 interior shelves. Thus, WISVA learns from 2,200 unique source-obstacle-receiver geometries, each presenting a distinct physical propagation ``puzzle.'' Shelf arrangements are varied across permutations to ensure the VAE encounters a statistically significant diversity of LOS and NLOS scenarios. This combinatorial diversity ensures the VAE learns the underlying physical manifold of 60~GHz propagation—specifically the relative source-obstacle geometry—rather than simply memorizing global floorplan templates. Consequently, the 2,200 diverse geometries provide a sufficiently dense sampling of incident angles and Fresnel zone clearances to enable true extrapolation, as validated in Section \ref{Sec_Extrapolation}.

For consistency, the AP is fixed at 15~meters elevation with an omnidirectional pattern, and the simulation resolution is 0.11~meters. Each floorplan point corresponds to a heatmap pixel representing the local SINR. The model predicts distributions for new layouts via interpolation—analyzing low-dimensional signal data—and extrapolation, predicting behavior in unexplored areas based on learned patterns. Finally, physical attributes including Euclidean distance, shelf permittivity, and AP coordinates are converted into tensors. Incorporating these properties allows the VAE to accurately estimate SINR heatmaps for novel configurations not explicitly represented in the training set.

\subsection{Input and Output Tensors in WISVA: Understanding the Warehouse Environment}

\begin{figure}[htbp!]
\centering 
\subfigure[Euclidean distance tensor]{\label{fig:TrainingTensors:Euclidean distance tensor}\includegraphics[height=34mm]{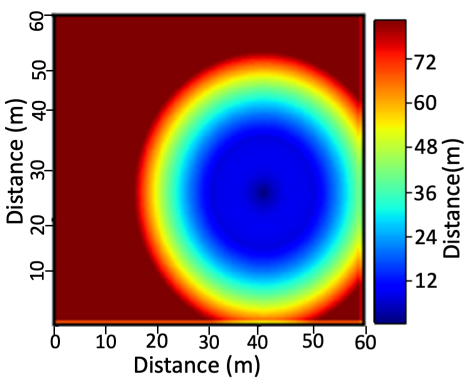}}
\hspace{1mm}
\subfigure[Permittivity tensor]{\label{fig:TrainingTensors:Permittivity tensor}\includegraphics[height=34mm]{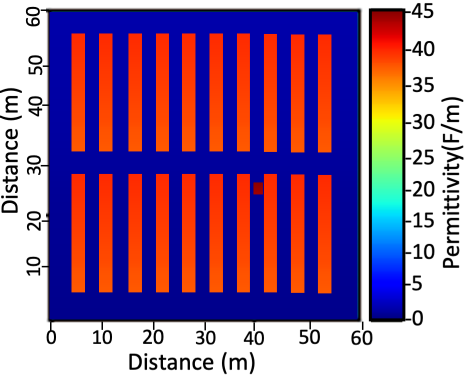}}
\hspace{1mm}
\subfigure[AP location tensor]{\label{fig:TrainingTensors:AP location tensor}\includegraphics[height=33mm]{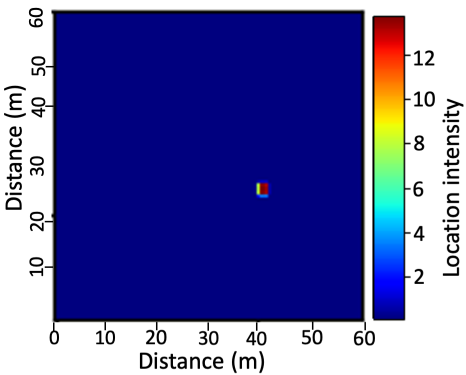}}
\hspace{1mm}
\subfigure[SINR from ns-3]{\label{fig:TrainingTensors:SINR from ns-3} \includegraphics[height=33mm]{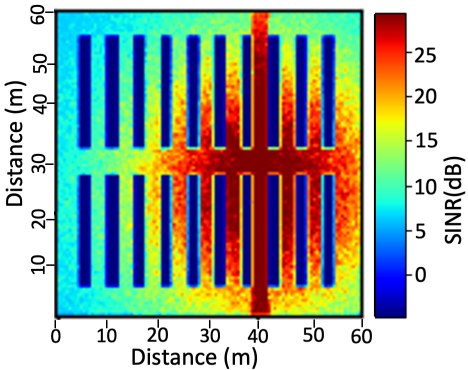}}
\caption{Training dataset: (a) Euclidean Distance, (b) Permittivity, (c) AP Location, and (d) ns-3 Simulated SINR.} 
\label{fig:TrainingTensors}
\end{figure}

The WISVA model's effectiveness hinges on a training dataset rich in diverse indoor physical properties \cite{EM_DeepRay}. Key tensors are illustrated in Figure~\ref{fig:TrainingTensors}.

The Euclidean Distance Tensor ($D(x, y)$) captures distance-dependent signal attenuation by calculating the physical separation from the AP at every point $(x, y)$:
\begin{equation}
D(x, y) = \sqrt{(x - x_{AP})^2 + (y - y_{AP})^2}
\label{eq:Distance_tensor} 
\end{equation}

The Effective Permittivity Tensor ($\epsilon_{eff}(x, y)$) accounts for structural obstructions by mapping material properties to their attenuation characteristics \cite{b10}. Rather than representing raw material constants, this tensor utilizes Effective Dielectric Constants to link physical obstacles to their 60~GHz propagation behavior:

\begin{equation}
\epsilon_{eff}(x, y) =
\begin{cases}
\epsilon_{air} = 1.0 & \text{if air/empty space} \\[-5pt] 
\epsilon_{shelf} = 35.5 & \text{if a shelf} 
\end{cases}
\label{eq:Permittivity_tensor} 
\end{equation}

By assigning a magnitude of $\epsilon_{shelf} = 35.5$, we map the empirical penetration loss of 35.5~dB for 60~GHz composite and metallic obstructions \cite{b10} directly into the permittivity domain. This contrast enables the VAE to accurately detect shadow boundaries in NLOS scenarios, as 60~GHz propagation is dominated by shadowing and penetration losses rather than multi-path diffraction.

The AP Location Tensor ($L_{\text{AP}}(x, y)$) explicitly encodes the transmitter coordinates to assist the model in identifying Line-of-Sight (LOS) propagation paths:
\begin{equation}
L_{\text{AP}}(x, y) =
\begin{cases}
\delta(x - x_{\text{AP}},\, y - y_{\text{AP}}), & \text{at } (x_{\text{AP}}, y_{\text{AP}}) \\
0, & \text{otherwise}
\end{cases}
\label{eq:AP_location_tensor}
\end{equation}

The Output Tensor is the ground-truth SINR heatmap $S_{dB}(x, y)$ simulated by \text{ns-3}, representing the signal-to-interference-plus-noise ratio in decibels:
\begin{equation}
S_{dB}(x, y) = 10 \log_{10} \left( \frac{P_{signal}(x, y)}{P_{interference}(x, y) + P_{noise}} \right)
\label{eq:SINR_dB}
\end{equation}

\subsubsection{Stochastic Modeling and Granularity}
To align tensors with physical realities, we address the relationship between discretization and channel stochasticity. While \text{ns-3} ground truth is generated at 0.11~m resolution, WISVA utilizes a $152 \times 152$ grid (~0.4~m per pixel) to align with slow-fading parameters. Since 60~GHz deterministic nulls ($<0.02$~m) are typically mitigated by hardware beamforming, WISVA targets the spatially-stable SINR envelope. This prevents overfitting to transient multipath artifacts and ensures reliable coverage predictions for industrial network planning.


\section{Proposed WISVA Model}

This section delves into the core of WISVA: the VAE model, exploring its workflow, the encoder and decoder layers, and how the architecture achieves accurate SINR prediction in automated warehouses.

\subsection{WISVA Encoder-Decoder Architecture: Capturing the Physics informed Characteristics of Warehouses}

The WISVA model adopts a VAE architecture, which learns compressed representations of complex data. This architecture consists of an Encoder and a Decoder.

\begin{figure}[htbp!]
\centering 
\subfigure[VAE model Encoder Structure]{\label{fig:EncoderLayer1}\includegraphics[height=75mm]{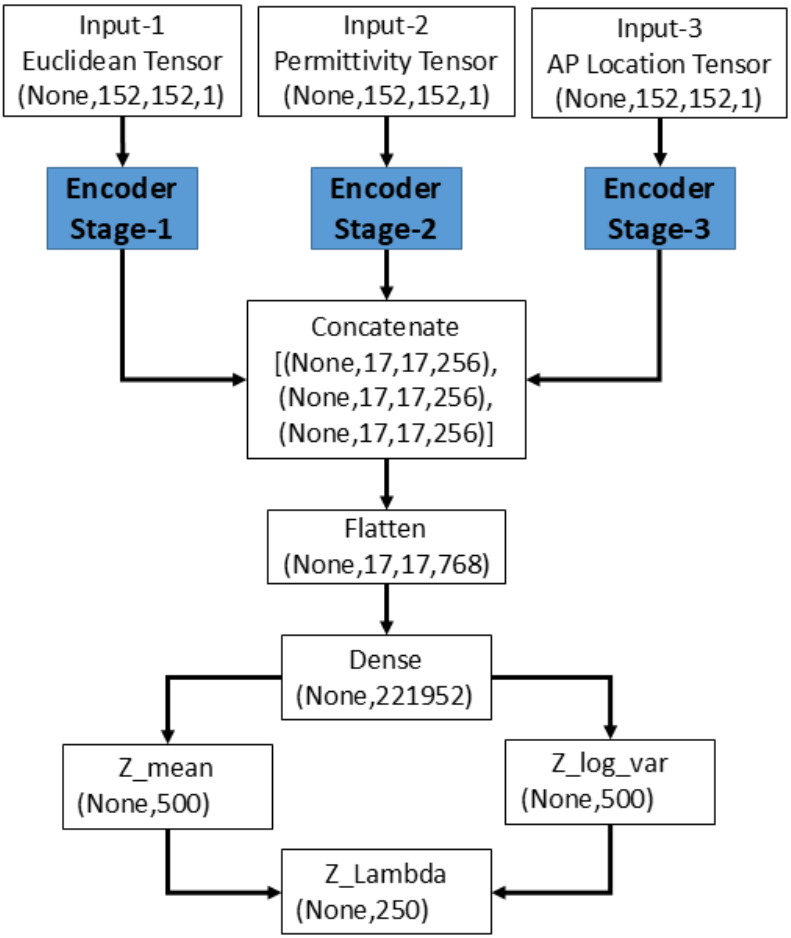}}
\caption{Detailed WISVA three-Branch Encoder Architecture.} 
\label{fig:EncoderLayers}
\end{figure}

The Encoder, depicted in Figure \ref{fig:EncoderLayers}, takes three input tensors—Euclidean distance ($E$), permittivity ($P$), and AP location ($A$)—representing the physical characteristics of a warehouse configuration.
\begin{itemize}
    \item Euclidean distance ($E$): A 2D tensor where $E_{(i,j)}$ is the distance between points $i$ and $j$.
    \item Permittivity ($P$): A 3D tensor where $P_{(i,j,k)}$ is the permittivity value at location $(i,j)$ on layer $k$.
    \item AP location ($A$): A 2D tensor where $A_{(i,j)}$ indicates the coordinates of the AP.
\end{itemize}
The encoder processes each tensor independently using Convolutional Neural Network (CNN) layers. These processed features are then concatenated and compressed into the latent space vector $Z$, achieving dimensionality reduction and efficiently capturing the most significant features that influence SINR behavior.

The Decoder reconstructs the SINR heatmap $\hat{H}$ from the latent vector $Z$. To eliminate non-physical checkerboard artifacts common in strided transposed convolutions, we utilize a series of nearest-neighbor upsampling layers followed by $3 \times 3$ convolutions. This architecture ensures spatially smooth signal gradients and physically consistent transitions at obstacle boundaries. The ReLU activation is applied in the normalized output space of the decoder; after inverse normalization, the reconstructed SINR maps span the full physical range of $-5$ to $30$~dB, including negative values corresponding to deep NLOS regions.

The optimal values for the architectural hyperparameters, including kernel sizes, filters, and pooling windows, were determined through a structured grid search aimed at minimizing the validation MAE. Specifically, we evaluated kernel sizes $k \in \{3, 5, 7\}$, while $7 \times 7$ kernels captured broader spatial context, $3 \times 3$ kernels were ultimately selected for the primary layers to maintain a lower parameter count and prevent over-smoothing of sharp SINR transitions at shelf edges. The filter depth was progressively doubled from 64 to 256 to allow the model to learn increasingly abstract physical features. Max-pooling with a $2 \times 2$ window was chosen to ensure spatial invariance, essential for accurately mapping AP signal footprints across varying warehouse coordinates.

The WISVA model undergoes intensive training iteratively to refine internal parameters to minimize a combination of two key loss functions:

The MAE Loss measures the pixel-wise difference between the reconstructed SINR heatmap $\mathbf{\hat{X}}$ and the ground truth $\mathbf{X}$ \cite{MAE_1}. By minimizing MAE loss, WISVA ensures the reconstructed heatmap closely resembles the real signal distribution:
\begin{equation}
\text{MAE} = \frac{1}{n} \sum_{i=1}^{n} \left| \mathbf{X}_i - \mathbf{\hat{X}}_i \right|
\label{eq:MAE}
\end{equation}
where $n$ is the total number of pixels, $\mathbf{X}_i$ is the actual SINR value, and $\mathbf{\hat{X}}_i$ is the reconstructed SINR value at pixel $i$.

The Kullback-Leibler (KL) Divergence shapes the distribution of the latent space representations learned by the encoder, measuring the discrepancy between the latent space distribution $q(\mathbf{z}|\mathbf{X})$ and a predefined prior distribution $p(\mathbf{z})$, typically a standard normal distribution $\mathcal{N}(0, \mathbf{I})$ \cite{variational_autoencoding}. Minimizing KL divergence encourages a well-structured and smoothly distributed latent space, fostering interpretability and facilitating efficient sampling:
\begin{equation}
D_{KL}(q(\mathbf{z}|\mathbf{X}) \parallel p(\mathbf{z})) = \int q(\mathbf{z}|\mathbf{X}) \log \frac{q(\mathbf{z}|\mathbf{X})}{p(\mathbf{z})} d\mathbf{z}
\label{eq:KL_divergence}
\end{equation}
This synergistic approach—MAE for fidelity and KL divergence for latent space structure—empowers WISVA to achieve accurate SINR prediction and generalizability.

The VAE was optimized using the Adam algorithm with an initial learning rate of $10^{-4}$. To ensure the architectural robustness of WISVA, a systematic grid search was conducted over the hyperparameter space $\mathcal{S} = \{ (\text{Batch}, \beta) \mid \text{Batch} \in \{2, 4, 8, 16, 32\}, \beta \in \{10^{-3}, 10^{-2}, 10^{-1}, 0.5, 1.0, 1.5, 2.0\} \}$. Empirical results demonstrated that the configuration of $\text{Batch}=16$ and $\beta=0.1$ constitutes an optimal Pareto-front, effectively balancing the high-fidelity reconstruction required for sub-decibel SINR precision with the latent space regularization necessary for consistent generalization. While larger batch sizes (e.g., $B=32$) maximize GPU memory utilization, they were found to suffer from the "large-batch generalization gap," where decreased stochastic gradient noise leads to convergence in sharp local minima, increasing the validation MAE. Training was executed for 1000 epochs on an NVIDIA V100 GPU, incorporating an early stopping mechanism to mitigate overfitting while preserving the structural integrity of the predicted heatmaps.


\subsection{WISVA Model Architecture and Training Data}

WISVA's training relies on simulated data generated by the ns-3 5G Lena module, which creates realistic SINR heatmaps serving as the ground truth. The training dataset generation employs a meticulous approach: Multiple APs are strategically positioned within a virtual warehouse environment using a grid pattern with a spacing of $5 \times 5$ meters. This systematic placement, repeated across various warehouse layouts, ensures comprehensive coverage and introduces the VAE to a broad spectrum of potential configurations, enhancing its ability to generalize.

The ns-3 5G Lena simulator incorporates advanced features, including realistic signal propagation that accounts for reflective and scattering effects, and uses discrete-event techniques to calculate path loss (PL) or SINR for both LOS and NLOS propagation paths \cite{Evaluation_60GHz, AutomatedWarehouse}.

The simulated data is preprocessed and transformed into a multi-channel input tensor $X$. These channels encode crucial information that influences signal propagation: Shelf Configuration, Horizontal Distance Map, LOS/NLOS Presence, Nearest Shelf Distance, and Permittivity Channel Maps. The inclusion of permittivity variations and the ability to handle non-rectangular warehouse layouts set WISVA apart from existing approaches (e.g., \cite{EM_DeepRay}) that often rely on simpler planar physics models, allowing WISVA to capture more realistic signal propagation effects.

The simulated data is preprocessed and transformed into a multi-channel input tensor $X$. These channels encode crucial information influencing signal propagation: Shelf Configuration, Horizontal Distance Map, LOS/NLOS Presence, Nearest Shelf Distance, and Permittivity Channel Maps. The resulting SINR values in the dataset and model outputs span a range of $[-5, 30]$~dB. This range is explicitly maintained through normalization rather than clipping, ensuring that negative SINR values and deep-fade regions critical for automated vehicle connectivity are fully preserved. The inclusion of permittivity variations and the ability to handle non-rectangular warehouse layouts set WISVA apart from existing approaches (e.g., \cite{EM_DeepRay}) that often rely on simpler planar physics models.

To expedite initial training and reduce computational requirements, the VAE is first trained on a grayscale version of the input tensors to establish a foundational understanding, before transitioning to the full multi-channel data. The WISVA model training process accommodates various warehouse configurations, resizing them to $152 \times 152$, ensuring the model's generalization across diverse indoor setups. This adaptability is crucial since warehouses often do not conform to perfect square or rectangular shapes.

\subsection{Latent Space Analysis of WISVA}
\label{sec:Chapter-4_3_4}

The generative strength of the WISVA framework is fundamentally derived from the organization of its Latent Space Representation (LSR). For a VAE to function as a robust generative model, its latent space must exhibit both continuity and completeness.

\subsubsection{The Role of KL Divergence in Latent Organization}
\label{sec:Chapter-4_3_4_1_KL_Role}

This structural integrity is achieved through the Kullback-Leibler (KL) Divergence term in the loss function. Without this term, the encoder tends to map warehouse configurations to isolated, deterministic points in the latent space with near-zero variance. Such a scattered, disconnected space is unsuitable for generation because sampling from the prior would likely yield incoherent results.

The KL term acts as a regularizer, forcing the learned distributions $q(\mathbf{z}|\mathbf{X})$ to align with a standard normal prior $\mathcal{N}(0, 1)$ \cite{variational_autoencoding}. Before regularization, we observe tightly peaked distributions—similar to a standard Autoencoder (AE)—which the model uses to achieve maximum reconstruction. While the AE focuses on compressing data into the densest possible points, the VAE's KL term ensures the LSR is ``cleaned'' and continuous. This enables the model to reliably sample any point in the latent space to generate a valid SINR map, effectively preventing ``hallucinations''—outputs that do not correspond to any valid warehouse configuration—which occur when an AE transitions between isolated clusters.

\subsubsection{Analysis and Visualization of the Latent Space}
\label{sec:Chapter-4_3_4_2_Analysis_Visualization}

\begin{figure}[htbp!]
    \centering
    \includegraphics[width=0.8\linewidth]{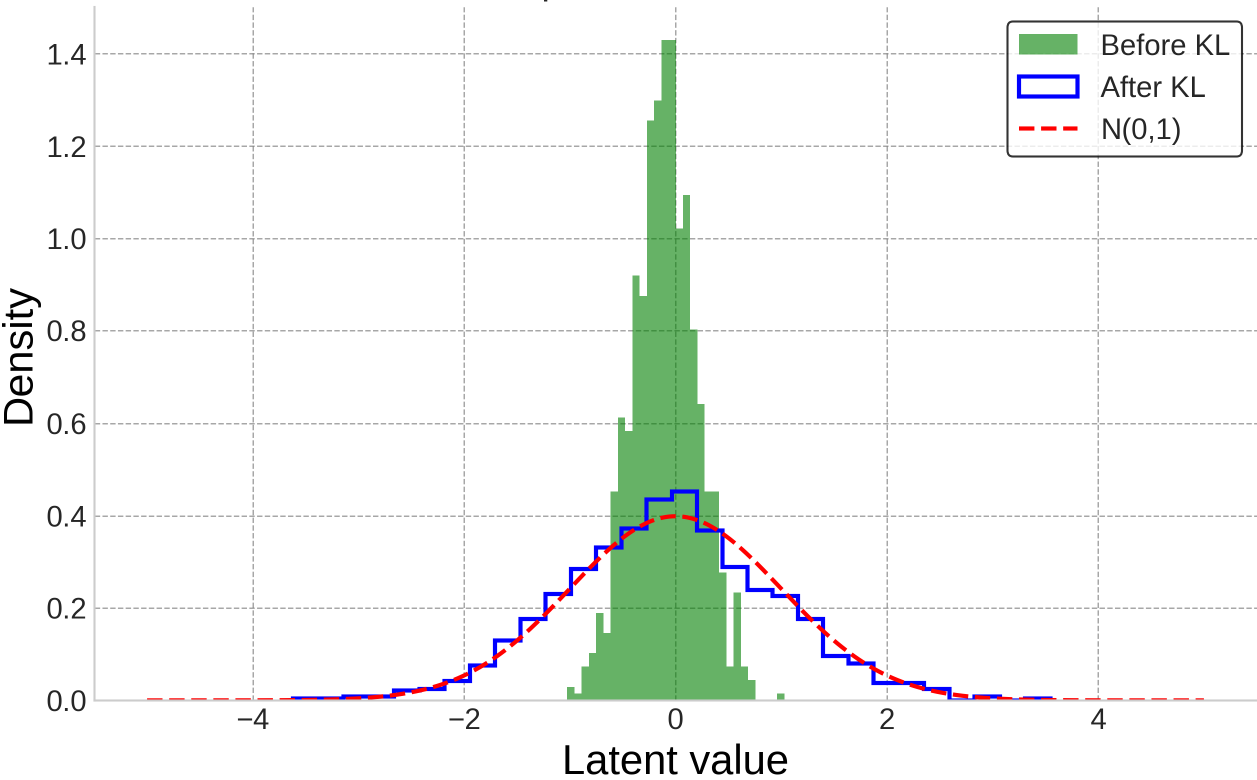} 
    \caption{Distribution analysis for latent dimension 243, comparing unregularized (green) vs. KL-regularized (blue) states against the $\mathcal{N}(0,1)$ prior (red dashed).}
    \label{fig:latent_dim_243}
\end{figure}

To illustrate the effect of KL divergence on individual latent dimensions, Figure \ref{fig:latent_dim_243} presents an analysis for a specific dimension (Dimension 243) from the WISVA VAE.

\begin{figure}[htbp!]
    \centering
    \includegraphics[width=0.7\linewidth]{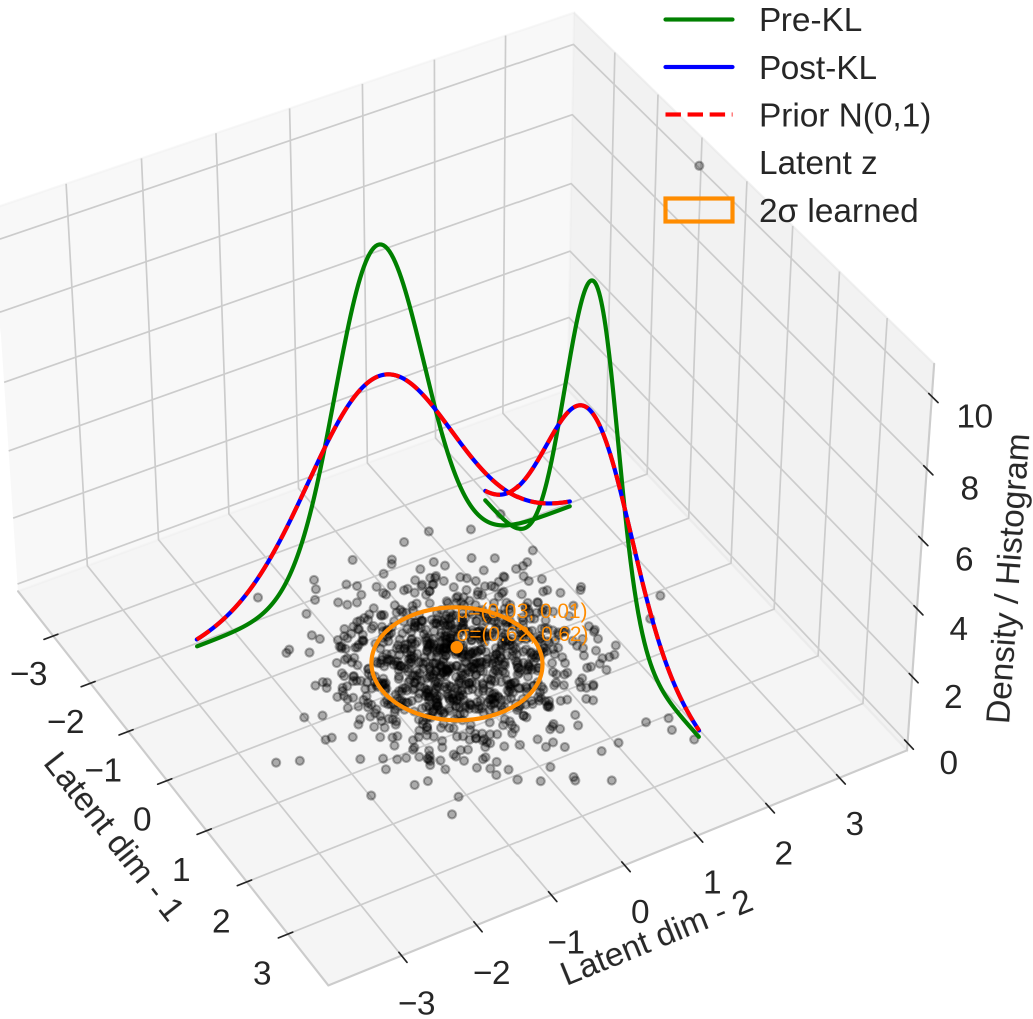}
    \caption{Conceptual VAE Latent Space in 3D with Pre- and Post-KL Marginals. Gray dots represent individual latent vectors $\mathbf{z}$.}
    \label{fig:vae_latent_3d}
\end{figure}

In Figure \ref{fig:latent_dim_243}, the ``Before KL'' histogram visualizes the natural tendency of the VAE to place latent values for this dimension. The notable overlap between the ``After KL'' histogram and the theoretical $\mathcal{N}(0,1)$ PDF confirms that the KL divergence term successfully guided this latent dimension toward the desired prior, indicating a well-regularized and continuous representation. The impact on the overall structure is conceptually visualized in 3D in Figure \ref{fig:vae_latent_3d}. Without KL regularization, individual latent distributions could be scattered and non-continuous. The KL divergence constrains these to align with a simple prior, forcing the latent vectors $\mathbf{z}$ to cluster meaningfully around the origin, creating a dense and navigable space suitable for drawing new, coherent samples.

\subsubsection{t-SNE Visualization and Channel State Analysis}
\label{sec:Chapter-4_3_4_3_tSNE_Visualization}

\begin{figure}[htbp!]
    \centering
    \subfigure[LSR Pre-KL (AE-like)]{\label{fig:LSR_Pre}\includegraphics[width=0.7\linewidth]{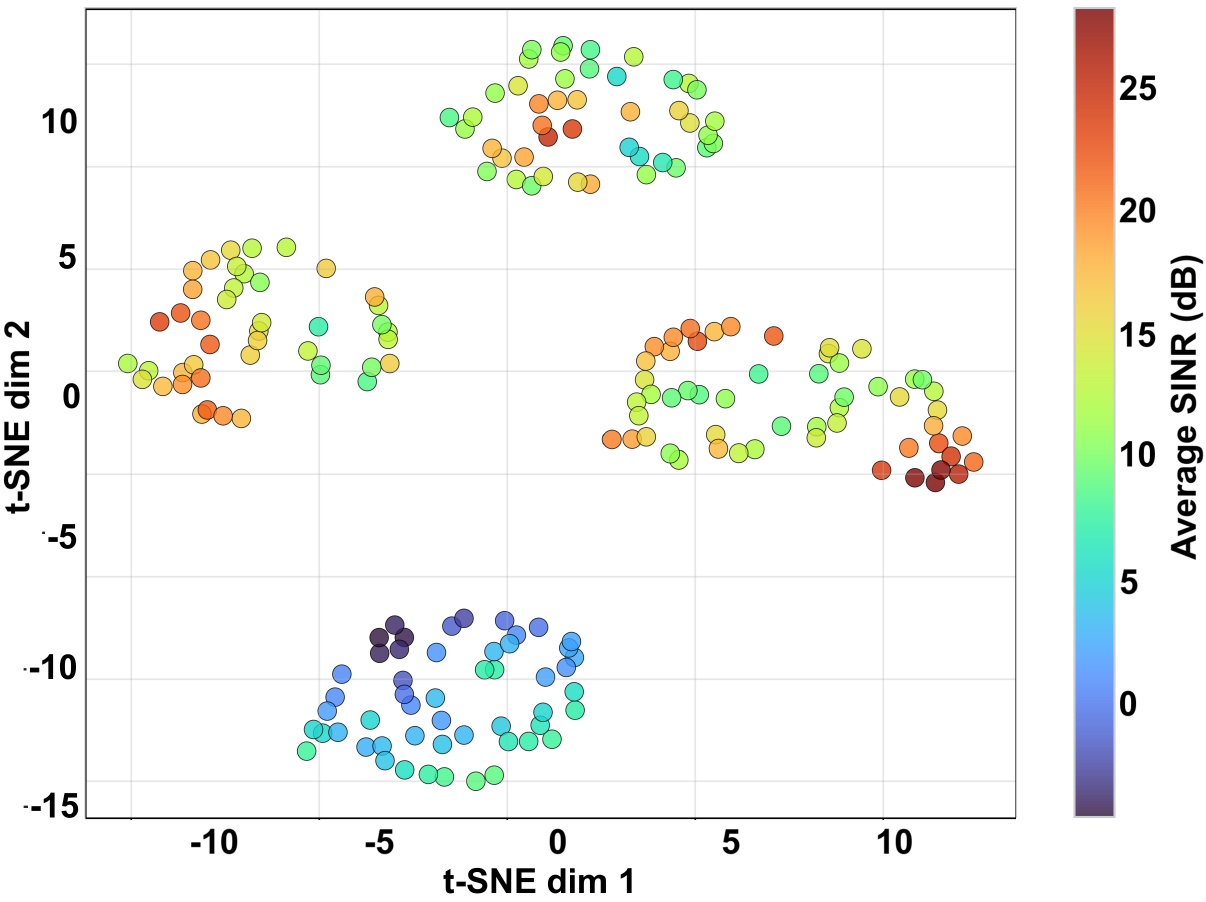}}
    \hfill
    \subfigure[LSR Post-KL (WISVA)]{\label{fig:LSR_Post}\includegraphics[width=0.7\linewidth]{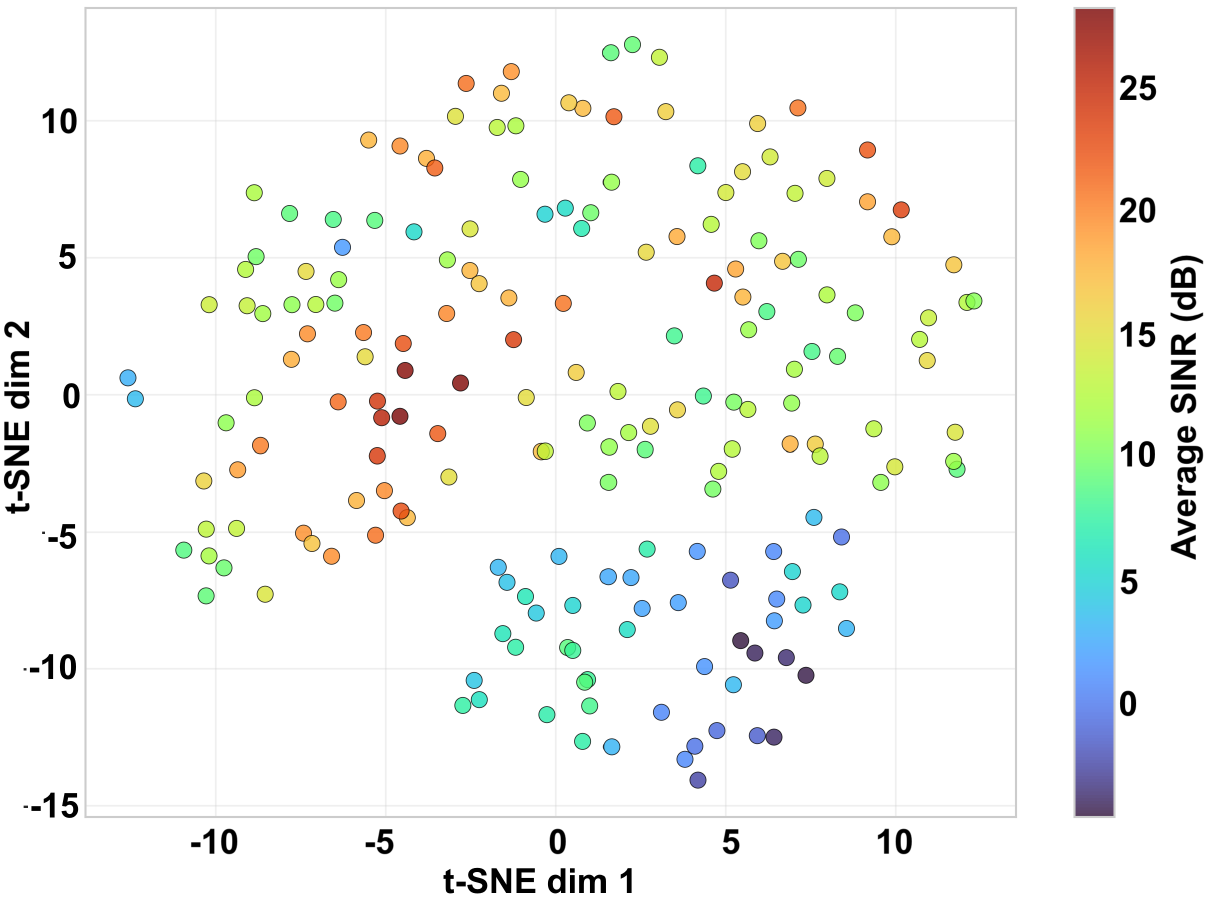}}
    \caption{t-SNE Visualization Comparison: (a) The pre-KL latent space shows four separate clusters representing dominant channel states. (b) The post-KL WISVA latent space forms a unified, continuous cloud enabling smooth interpolation.}
    \label{fig:tSNE_Comparison}
\end{figure}

To visualize the complex organization of the 256D WISVA latent space, we employ t-SNE \cite{tsne_citation}. Figure \ref{fig:tSNE_Comparison} presents a t-SNE projection of the latent vectors $\mathbf{z}$, colored by the average SINR (dB) of each configuration. The plot reveals a continuous distribution of latent points, characterized by smooth transitions in average SINR values across the 2D plane. The absence of distinct, sharply separated clusters indicates that WISVA has successfully learned a continuous and high-complexity representation that mirrors the nuances of real-world radio propagation.

\begin{figure}[htbp!]
    \centering
    \includegraphics[width=0.6\linewidth]{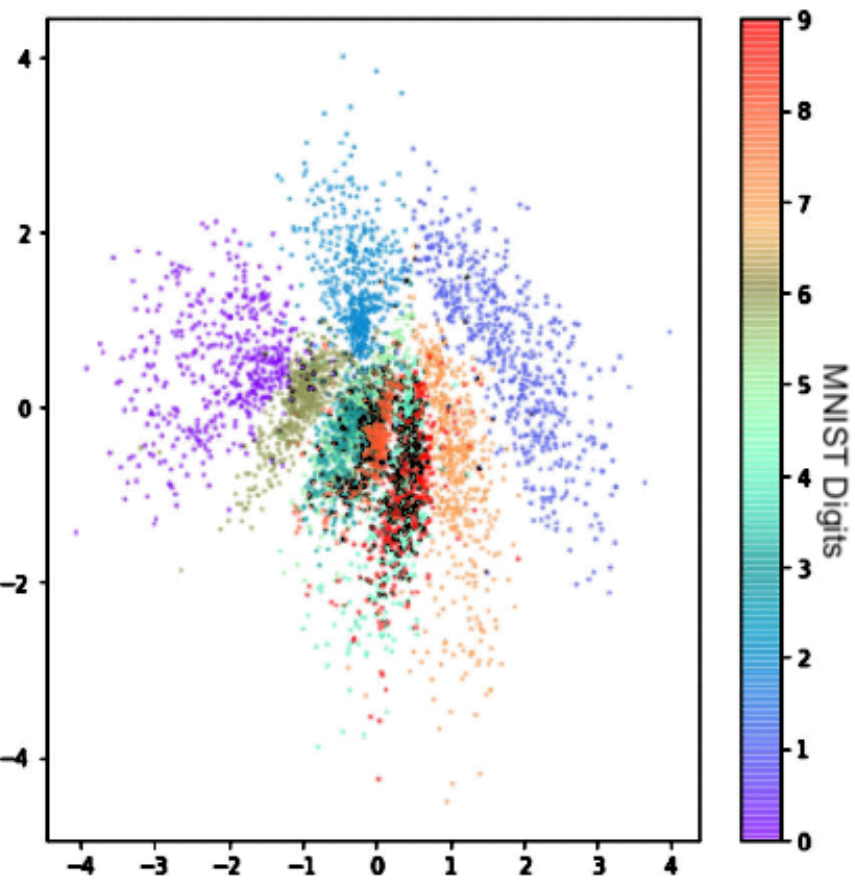} 
    \caption{t-SNE Visualization of a VAE Latent Space Trained on MNIST. Distinct, well-separated clusters corresponding to each digit highlight the categorical nature of the data.}
    \label{fig:MNIST_tSNE}
\end{figure}

As illustrated in Figure \ref{fig:tSNE_Comparison}, the pre-KL space exhibits four distinct, separate clusters. These represent the four dominant channel states the model found easiest to compress: strong Line-of-Sight (LOS) conditions near the AP, full Non-Line-of-Sight (NLOS) or total blockage regions, LOS conditions within warehouse aisles, and the transitionary edge of the coverage zone. For comparative context, the t-SNE visualization is benchmarked against the MNIST dataset \cite{mnist_citation} in Figure \ref{fig:MNIST_tSNE}. Unlike the well-separated clusters corresponding to each digit in MNIST, WISVA's fluid structure underscores the high complexity of modeling continuous wireless propagation phenomena compared to discrete classification, validating its capacity to represent complex radio channels effectively.

\subsubsection{LSR: Disentanglement, Independence, and Dimensionality}
\label{sec:Chapter-5_3_4_Disentanglement}

\begin{figure}[htbp!]
    \centering
    \includegraphics[width=0.8\linewidth]{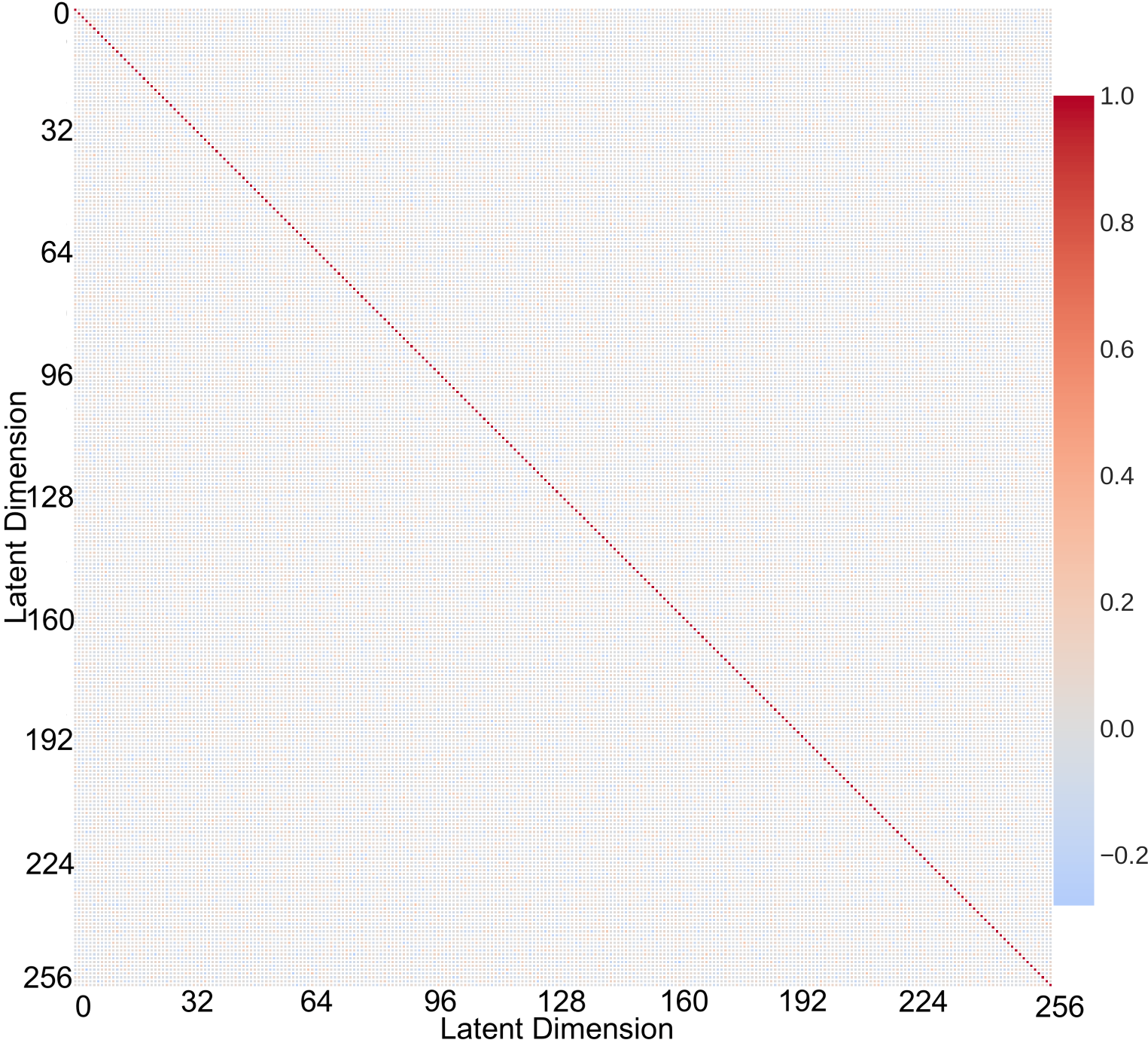}
    \caption{Pearson correlation heatmap of all 256 latent dimensions. The nearly zero off-diagonal values confirm a highly disentangled representation.}
    \label{fig:Pearson_Heatmap}
\end{figure}

A critical property of a robust VAE is disentanglement, where each of the 256 latent dimensions represents an independent factor of variation. This independence ensures the latent space is interpretable and suitable for autonomous optimization tasks. To quantify this, we performed a Pearson Correlation Analysis across all 256 dimensions. As shown in Figure \ref{fig:Pearson_Heatmap}, the off-diagonal elements are essentially zero, resulting in a mean absolute correlation of only $0.038$. This linear independence confirms that the VAE successfully disentangled the high-dimensional RF channel characteristics.

\begin{figure}[htbp!]
    \centering
    \includegraphics[width=0.7\linewidth]{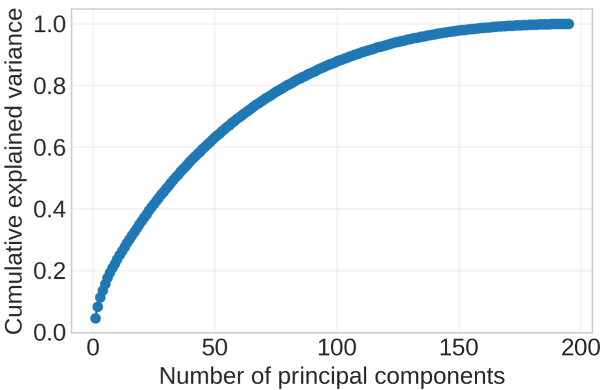}
    \caption{Cumulative Explained Variance by Principal Components.}
    \label{fig:PCA_Curve}
\end{figure}

To justify the choice of a 256-dimensional space, we utilized Principal Component Analysis (PCA) to evaluate the intrinsic complexity of the smart warehouse channel. The smooth, gradual Cumulative Explained Variance curve in Figure \ref{fig:PCA_Curve} reveals that variance is spread evenly across many dimensions. This confirms that modelling the RF channel is highly complex; lower-dimensional spaces (e.g., $D=64$) were found to produce blurred SINR maps that failed to capture sharp shadow boundaries. The choice of $D=256$ serves as an optimization trade-off, ensuring architectural simplicity and computational efficiency on GPUs.

To justify the choice of a 256-dimensional space, we utilized Principal Component Analysis (PCA) to evaluate the intrinsic complexity of the smart warehouse channel. The smooth, gradual Cumulative Explained Variance curve in Figure \ref{fig:PCA_Curve} reveals that variance is spread evenly across many dimensions. Empirical ablation of the latent bottleneck showed that lower-dimensional spaces, specifically $Z=64$, produced blurred SINR maps with a significant MAE increase of $+0.21$~dB compared to the baseline. While $Z=128$ improved boundary definition, $Z=256$ was found to be the optimal convergence point, as $Z=512$ offered no statistically significant MAE reduction. Thus, $D=256$ serves as the ideal optimization trade-off between reconstruction fidelity and computational efficiency.

\section{WISVA Model Evaluation}

In this section, we juxtapose the SINR predictions of the WISVA model with simulated data generated using the \text{ns-3} 5G Lena module. We analyze the model across various test cases, including performance benchmarking, spatial field reconstruction, extrapolation, and low-data generalization.

\subsection{Performance Benchmarking on the Validation Dataset}

\begin{figure}[htbp!]
\centering
\subfigure[SINR from ns-3]{\label{fig:SINR_GT}\includegraphics[height=33mm]{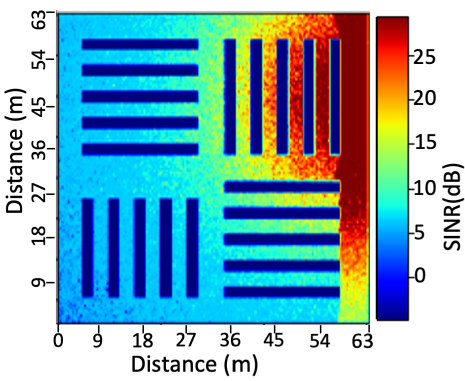}}
\hspace{1mm}
\subfigure[WISVA Predicted]{\label{fig:SINR_WISVA}\includegraphics[height=33mm]{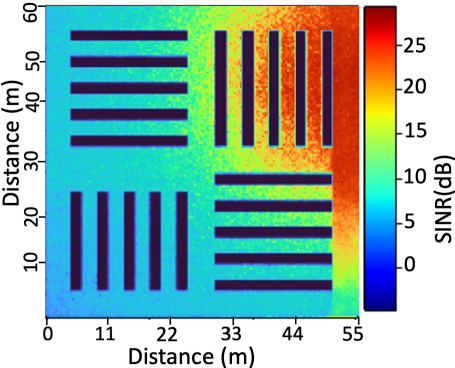}}
\hspace{1mm}
\subfigure[AE Predicted]{\label{fig:SINR_AE}\includegraphics[height=33mm]{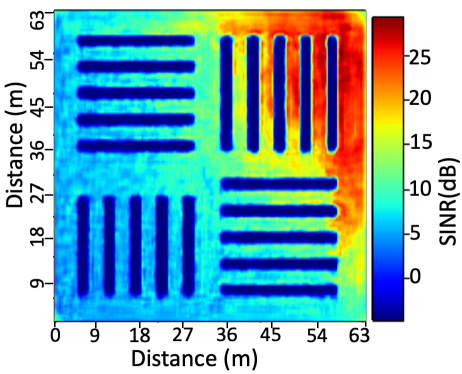}} 
\hspace{1mm}
\subfigure[RadioUNet Predicted]{\label{fig:SINR_RadioUNet}\includegraphics[height=33mm]{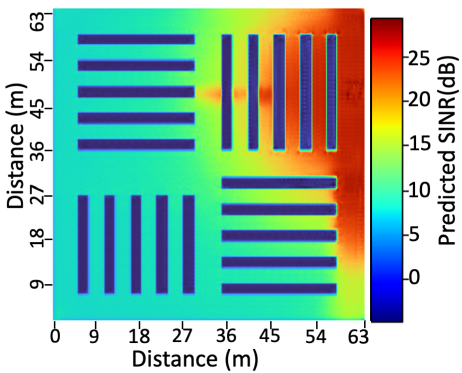}}    
\hspace{1mm}
\subfigure[EMDeep Predicted]{\label{fig:SINR_EMDeep}\includegraphics[height=33mm]{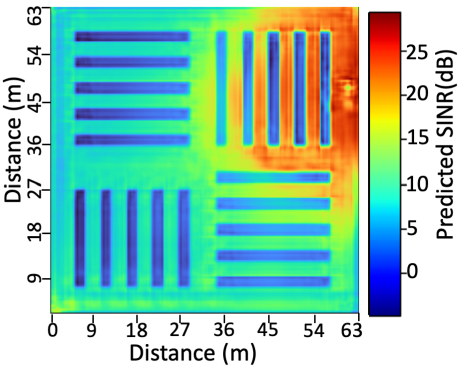}}     
\hspace{1mm}
\subfigure[NeRF² Predicted]{\label{fig:SINR_NeRF2}\includegraphics[height=33mm]{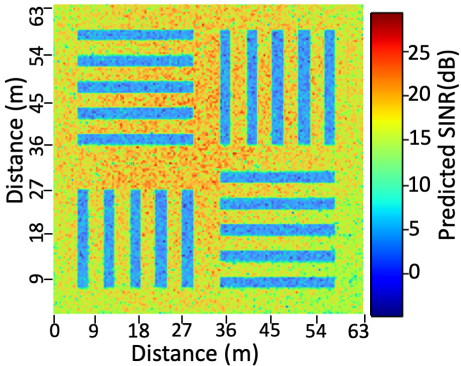}}        
\caption{Comparison of SINR prediction heatmaps: (a) Ground-truth ns-3, (b) WISVA, (c) Autoencoder (AE), (d) RadioUNet, (e) EMDeep, and (f) $\text{NeRF}^{2}$ models.}
\label{fig:SINR_Model_Comparison}
\end{figure}

We evaluate the WISVA model using a validation dataset, with a sample comparison presented in Figure \ref{fig:SINR_Model_Comparison}. The model accurately predicts the SINR heatmap for specified AP locations and warehouse configurations, achieving minimal reconstruction errors. Dataset diversity is ensured through dense spatial sampling of AP positions across multiple layouts, which serves as an implicit physical augmentation to prevent overfitting to geometric symmetries while preserving electromagnetic consistency.

To assess performance, we benchmarked WISVA against the AE \cite{Comparison_AE}, RadioUNet \cite{Comparison_RadioUNet}, and EM Deep Ray \cite{EM_DeepRay} models, with a qualitative comparison against coordinate-based $\text{NeRF}^{2}$ \cite{Comparison_NERF2}. As shown in Figure \ref{fig:SINR_Model_Comparison}, baseline models exhibited varying reconstruction loss, resulting in blurry or distorted outputs. While $\text{NeRF}^{2}$ provides a continuous field representation, its application to discretized SINR tensors in high-clutter environments yields significant artifacts compared to WISVA's latent-space reconstruction. By integrating physics-informed tensors, WISVA generates accurate heatmaps with lower cumulative reconstruction errors across all baselines. Notably, accuracy was highest in non-line-of-sight (NLOS) regions; despite transition zones between LOS and NLOS presenting the highest challenge, WISVA consistently yielded the lowest maximum average per-pixel error.

Quantitative metrics in Table \ref{tab:Table1} confirm WISVA's efficacy, demonstrating a Mean Absolute Error (MAE) of $0.33\text{ dB}$, outperforming $\text{RadioUNet}$ \cite{Comparison_RadioUNet}, EM Deep Ray \cite{EM_DeepRay}, $\text{NeRF}^{2}$ \cite{Comparison_NERF2}, AE \cite{Comparison_AE}, and the Multi-Wall Model (MWM) \cite{comparison_1}. Benchmarked on a Xeon Bronze 3204 CPU, a single \text{ns-3} simulation requires 15m:48s (Table \ref{tab:Table_SimTime}), while WISVA completes inference in $\approx$88~ms. This $>10^4\times$ speed-up enables real-time Digital Twin ``what-if'' analysis with sub-decibel precision. While not a substitute for site-specific empirical surveys, its utility as a high-speed surrogate for the 3GPP-compliant \text{ns-3} framework \cite{ns3_defense_1, ns3_defense_2} is clear where traditional ray-tracing is computationally prohibitive. Hardware feasibility for industrial edge deployment was also evaluated. Peak memory footprint at FP32 is $\approx$33.58~MB ($<$1\% of standard controller RAM), confirming the architecture's suitability for resource-constrained Fog nodes without specialized server-grade hardware.

\begin{table}[htbp!]
\centering
\caption{Quantitative Performance Benchmark: MAE Comparison}
\label{tab:Table1}
\renewcommand{\arraystretch}{1.2} 
\begin{tabular}{|l|c|}
\hline
\rowcolor[gray]{0.9} \textbf{Model} & \textbf{MAE [dB]} $\downarrow$ \\ \hline
\textbf{WISVA (Proposed)} & \textbf{0.329} \\ \hline
RadioUNet \cite{Comparison_RadioUNet} & 0.80 \\ \hline
EM Deep Ray \cite{EM_DeepRay} & 1.22 \\ \hline
Autoencoder (AE) \cite{Comparison_AE} & 4.16 \\ \hline
MWM \cite{comparison_1} & 4.65 \\ \hline
\end{tabular}
\end{table}

A sensitivity analysis across five independent training runs with randomized initializations confirmed high architectural stability. The resulting mean MAE was 0.334~dB ($\sigma = 0.012$~dB), with a 95\% confidence interval of $[0.317, 0.341]$~dB, validating that predictive precision is a robust characteristic of the VAE's latent organization rather than a stochastic artifact.

In conclusion, WISVA provides superior accuracy and detail in SINR heatmap reconstruction, particularly when compared to the AE model. Its millisecond-level inference speed (Table \ref{tab:Table_SimTime}) and high fidelity significantly improve the reliability of wireless network analysis in complex industrial settings.

\subsection{Spatial Field Reconstruction and Super-Resolution}

The design of WiGig infrastructure for real-time smart warehouse applications requires rapid, efficient simulation tools. However, \text{ns-3} often produces low-resolution, pixelated SINR heatmaps ($50 \times 50$ grid sampling) due to computational constraints (Table \ref{tab:Table_SimTime}). This study evaluates the WISVA framework's effectiveness in performing high-fidelity spatial field reconstruction, upsampling these heatmaps to $152 \times 152$ pixels to recover the underlying physical signal distribution.

\begin{figure}[t] 
\centering 
\subfigure[Noisy SINR heatmap]{\label{fig:Denoising_WISVA_AE_ns3_heatmap1}\includegraphics[width=0.48\linewidth]{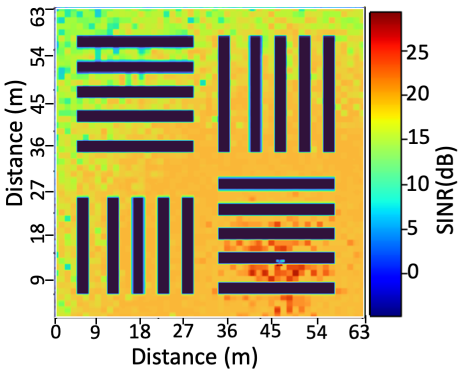}}       
\hfill 
\subfigure[WISVA SINR heatmap]{\label{fig:Denoising_WISVA_Heatmap}\includegraphics[width=0.48\linewidth]{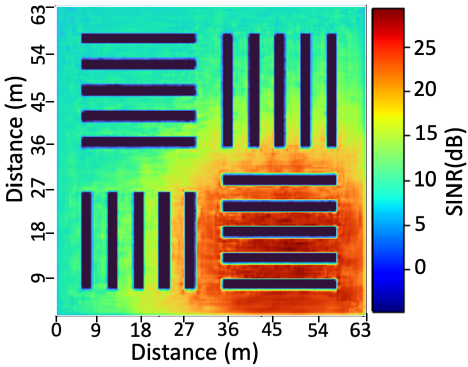}}      
\subfigure[WISVA Reconstruction Error]{\label{fig:Denoising_WISVA_Error}\includegraphics[width=0.48\linewidth]{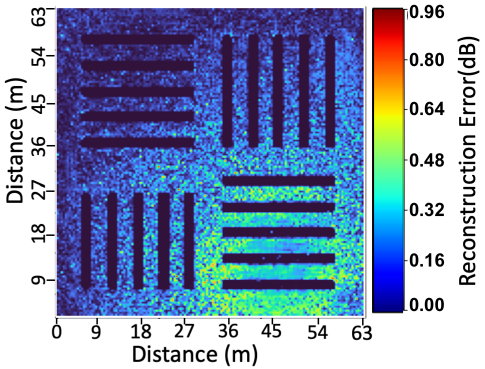}}      
\caption{Analysis of WISVA for spatial field reconstruction: (a) ns-3 generated noisy SINR heatmap, (b) WISVA Predicted Heatmap, and (c) Absolute Reconstruction Error.} 
\label{fig:WISVA_AE_Denoising}
\end{figure}

To address aliased and sparse SINR heatmaps resulting from prioritized computational efficiency in \text{ns-3}, the VAE-based WISVA model processes low-fidelity inputs (Figure \ref{fig:Denoising_WISVA_AE_ns3_heatmap1}) to reconstruct high-resolution representations (Figure \ref{fig:Denoising_WISVA_Heatmap}). This super-resolution process utilizes physics-informed knowledge of the warehouse layout to mitigate spatial aliasing and quantization artifacts while preserving critical signal strength. Precision is quantified via the MAE metric, with error heatmaps (Figure \ref{fig:Denoising_WISVA_Error}) showing the absolute difference between high-resolution predictions and low-fidelity inputs. Darker regions in these maps indicate high structural similarity, visually confirming that the model effectively recovers key signal characteristics while mitigating artifacts induced by sparse sampling.


The findings reveal that WISVA achieves lower maximum and average reconstruction errors, highlighting its superior ability to capture intricate data patterns while maintaining stable learning behavior across varying dataset sizes.

\subsection{Extrapolation to Unseen Layouts and AP Positions}
\label{Sec_Extrapolation}

To verify that WISVA learns generalizable physical laws rather than memorizing training templates, its extrapolation potential was rigorously tested by withholding quadrant IV entirely during training. This ``blind quadrant'' test is critical: because 60~GHz propagation is site-specific, quadrant IV presents novel source-obstacle geometries. While only nine discrete AP positions exist in this quadrant, evaluation occurs across 207,936 spatial points. The 5~m AP grid spacing ensures test samples are spatially decorrelated, validating the model’s ability to generalize complex EM interactions—such as corner diffraction and aisle-guided propagation—relative to the training manifold.

\begin{figure}[htbp!]
\centering 
\subfigure[Training-Testing quadrants]{\label{fig:Result3_Quadrants}\includegraphics[height=34mm]{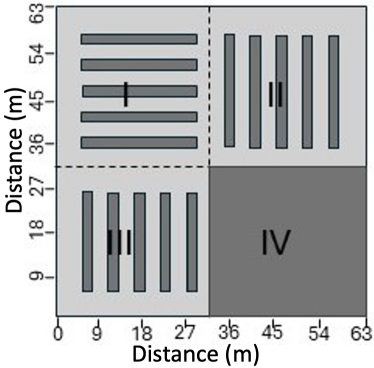}}     
\hspace{1mm}
\subfigure[SINR from ns-3]{\label{fig:Result3:SINRlabel} \includegraphics[height=34mm]{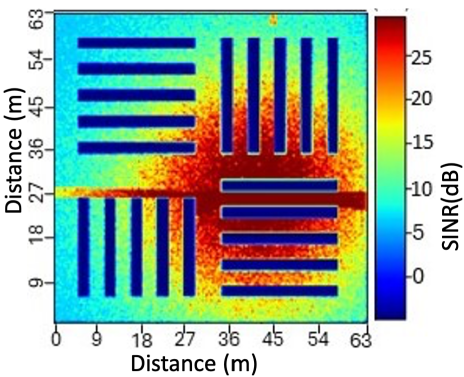}}    
\hspace{1mm}
\subfigure[WISVA Predicted]{\label{fig:Result3:SINRpredicted} \includegraphics[height=32mm]{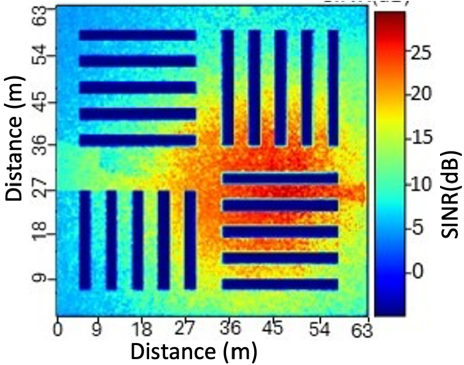}}   
\hspace{1mm}
\subfigure[WISVA Reconstruction error]{\label{fig:Result3:Reconstruction_error} \includegraphics[height=32mm]{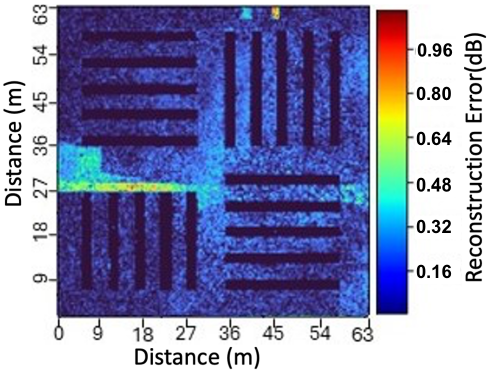}}   
\caption{Analysis of WISVA extrapolation: (a) dataset quadrant structure, (b) \text{ns-3} Ground-truth, (c) WISVA Prediction, and (d) Absolute Reconstruction Error.} 
\label{fig:Result3}
\end{figure}


The validation results (Figure~\ref{fig:Result3}) demonstrate WISVA’s ability to accurately predict SINR for unseen quadrants, effectively extending coverage beyond the trained environment. This is corroborated by the low error heatmap (Figure~\ref{fig:Result3:Reconstruction_error}). Observed error concentrations at LOS–NLOS boundaries stem from steep spatial SINR gradients near obstacle edges, where signal propagation shifts rapidly from direct transmission to diffraction-dominated paths. These transition zones involve complex phase variations that are inherently challenging to map within a compressed latent representation. We note that future work will explore boundary-aware loss functions to prioritize these high-gradient regions and further enhance transition precision.

WISVA’s successful extrapolation provides immense value for expanding smart warehouses by enabling proactive SINR optimization without costly retraining. While the model shows slight difficulty in frequent LOS-NLOS transition areas, augmenting training with spatially constrained environment signal heatmaps and refining the VAE’s latent space architecture will further improve boundary precision.

\subsection{Environment-aware Data Augmentation Analysis}
Data augmentation techniques are often adopted to evaluate image based deep learning and generative AI models. While techniques such as scaling, shearing, or elastic deformation are common in computer vision, they are unsuitable for wireless propagation modeling. Such transformations would alter the wavelength-to-obstacle ratio and break the geometric consistency of reflection and diffraction physics, corrupting the latent space of the physics-informed WISVA model. Consequently, evaluation was strictly limited to physics-preserving transformations: 180° rotation, horizontal flip, and translation (spatial shift). These ensure that the environment's scale and the AP’s point-source characteristics remain physically valid. As shown in Table~\ref{tab:Augmentation_Recall}, WISVA maintained an NLOS recall above 94\% across these transformations, ensuring the encoder learns generalizable EM structures rather than overfitting to specific shelf symmetries.

\begin{table}[htbp!]
\centering
\caption{Zero-Shot Robustness and Generalization Analysis}
\label{tab:Augmentation_Recall}
\renewcommand{\arraystretch}{1.2}
\resizebox{\columnwidth}{!}{
\begin{tabular}{|l|c|c|c|c|}
\hline
\rowcolor[gray]{0.9} \textbf{Augmentation Type} & \textbf{Accuracy} & \textbf{NLOS Recall} & \textbf{LOS Recall} & \textbf{MAE [dB]} \\ \hline
Standard Baseline & 98.12\% & 98.48\% & 97.47\% & 0.329 \\ \hline
Horizontal Flip   & 94.67\% & 94.83\% & 94.39\% & 0.458 \\ \hline
Rotation (180°)   & 92.91\% & 94.02\% & 90.98\% & 0.542 \\ \hline
Spatial Shift     & 92.64\% & 94.05\% & 90.22\% & 0.591 \\ \hline
\end{tabular}%
}
\end{table}

\subsection{Low-Data Generalization Capability}

\begin{figure}[htbp!]
\centering 
\subfigure[SINR from ns-3]{\label{fig:TestCase2_SINRlabel} \includegraphics[height=33mm]{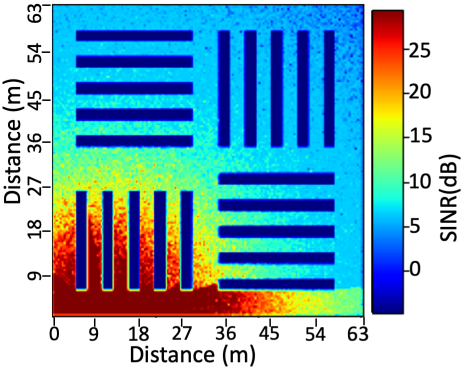}}      
\hspace{-1mm}
\subfigure[WISVA Predicted]{\label{fig:TestCase2_SINRpredicted} \includegraphics[height=33mm]{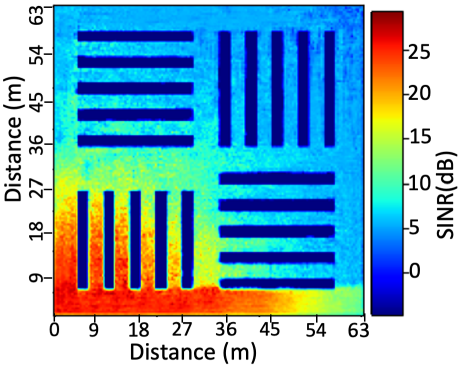}}     
\hspace{-1mm}
\subfigure[WISVA Reconstruction error]{\label{fig:TestCase2_Error} \includegraphics[height=33mm]{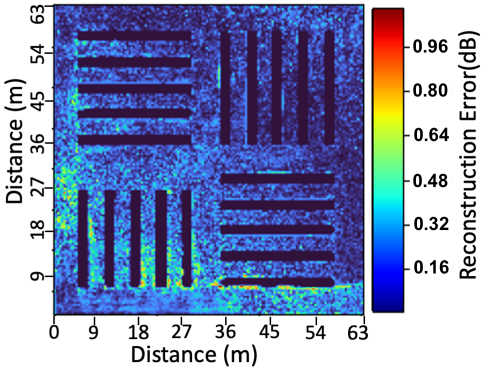}}      
\caption{Analysis of WISVA on minimally sampled dataset: (a) \text{ns-3} Ground-truth, (b) WISVA Prediction, and (c) Absolute Reconstruction Error.} 
\label{fig:TestCase2}
\end{figure}

To address real-world challenges in data availability, we investigated WISVA's performance under data-constrained conditions using a limited number of training samples from a novel warehouse setup. Figures \ref{fig:TestCase2_SINRpredicted} and \ref{fig:TestCase2_Error} illustrate the predicted SINR and error heatmaps derived from a random validation sample.

\begin{figure}[htbp!]
\centering
\includegraphics[width=0.90\linewidth]{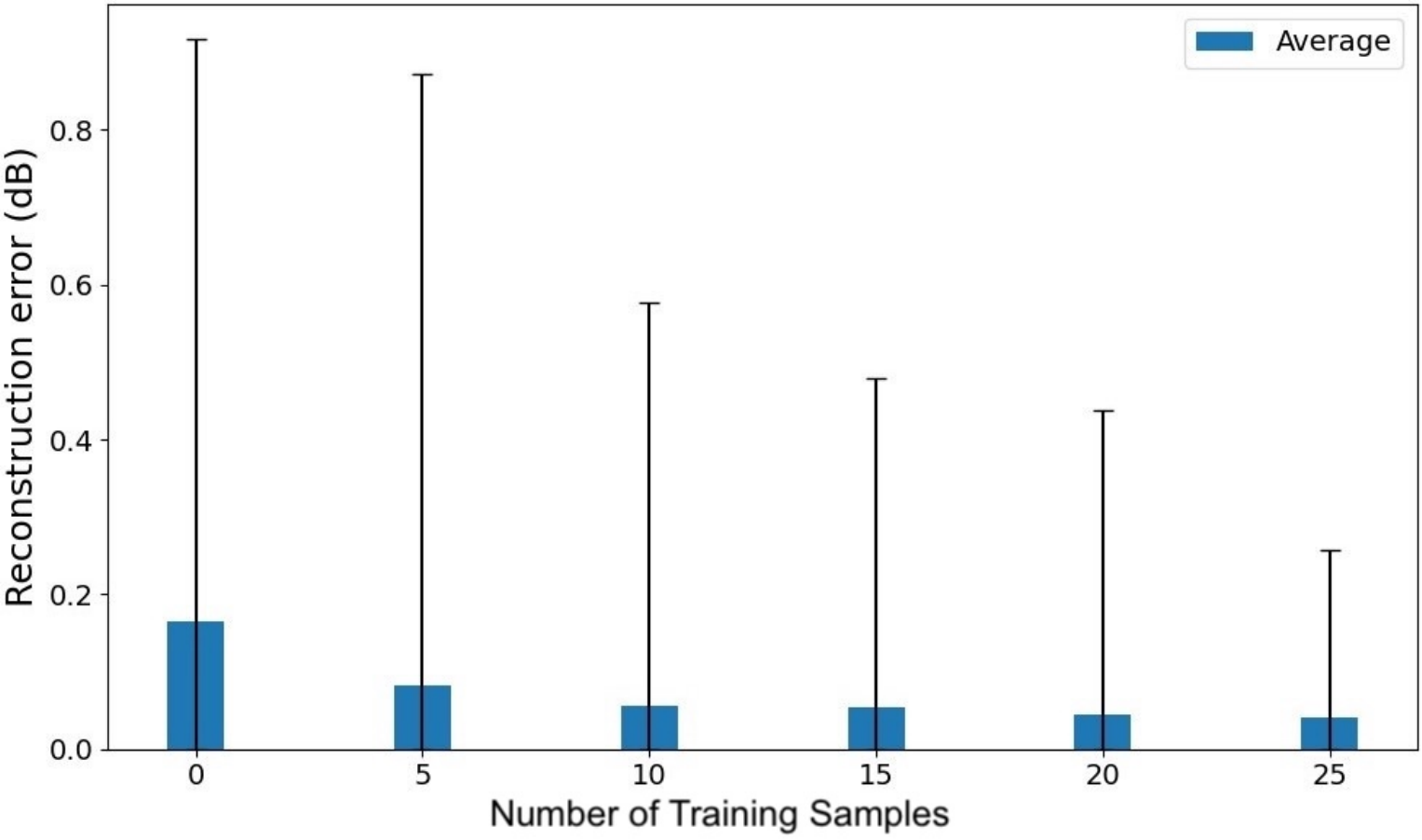}     
\caption{WISVA performance vs. reconstruction error on minimally sampled data. Whiskers represent the error range per pixel across samples.}
\label{fig:TestCase2_ErrorAnalysis}
\end{figure}

A high degree of similarity is observed between the \text{ns-3} simulated and WISVA-predicted heatmaps, with acceptable localized errors near the AP. Error analysis (Figure \ref{fig:TestCase2_ErrorAnalysis}) demonstrates a consistent decrease in both average and peak error as more samples are incorporated, confirming WISVA's ability to generalize effectively and elucidating the fundamental trade-off between training volume and prediction precision.

In contrast, the AE model exhibits substantially higher reconstruction losses, yielding blurry and distorted outputs. By leveraging the VAE framework and physics-informed tensors, WISVA generates more detailed heatmaps, particularly in challenging LOS-to-NLOS transitions. Quantitatively, WISVA achieves a maximum average per-pixel reconstruction error of $0.329\text{ dB}$, significantly outperforming the AE's $4.16\text{ dB}$. These findings underscore WISVA's robustness and potential as a high-fidelity tool for wireless network design in smart environments, even when operating with limited training data.


\section{Conclusion}

This work introduced WISVA, a VAE-based framework for accelerated surrogate modeling of 5G infrastructure in smart warehouses. By capturing intricate electromagnetic interactions, WISVA approximates 3GPP-compliant \text{ns-3} 5G LENA data with a mean absolute error of 0.329~dB at a fraction of traditional computational costs. This 0.329~dB figure represents a relative gain within the standardized \text{ns-3} framework; while the model effectively captures stochastic paths, future calibration against full-wave solvers is required to account for metallic specular reflections from industrial shelving.

Beyond accuracy, WISVA's millisecond-level inference and robust performance under data-constrained conditions make it highly suitable for real-time Digital Twin deployments and proactive network management. The integration of physics-informed tensors into the latent space enhances generalization across unseen layouts, providing critical insights for rapid network optimization. While currently a single-source propagation primitive, the nature of 60~GHz signals allows for future expansion into multi-AP interference modeling via linear superposition. Future research will focus on refining sharp LOS-NLOS transition modeling and extending the framework to diverse indoor industrial domains.

\section{Acknowledgment}
This work was supported by the Raymond Corporation University Research Program. All research was conducted using open-source simulation frameworks (ns-3), and the grantor had no role in the study design, data collection, analysis, or the decision to publish these findings. All conclusions reflect the independent scientific judgment of the authors.


{\small
\bibliographystyle{IEEEtran}
\bibliography{main}}

\begin{IEEEbiography}[{\includegraphics[width=1in,height=1.25in,clip,keepaspectratio]{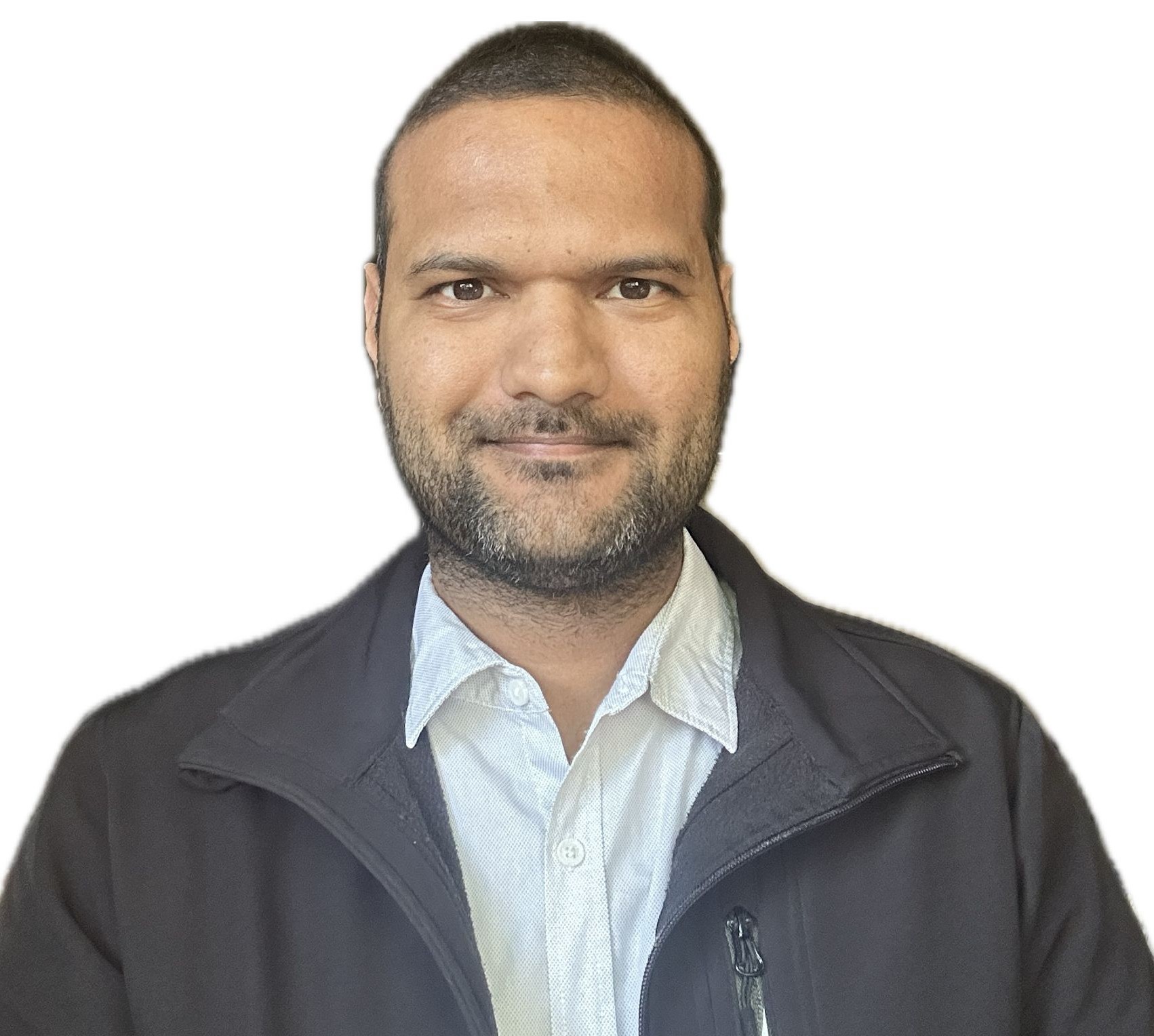}}]{Rahul Gulia} (S'20) received the M.S. degree in Electrical Engineering from the Rochester Institute of Technology, where he is currently pursuing the Ph.D. degree in Electrical and Computer Engineering. His research interests include AI-driven wireless communication systems, mmWave propagation modeling, and machine learning for industrial wireless networks. His work focuses on physics-informed deep learning and real-time radio environment prediction for Industry 4.0 applications.
\end{IEEEbiography}

\begin{IEEEbiography}[{\includegraphics[width=1in,height=1.25in,clip,keepaspectratio]{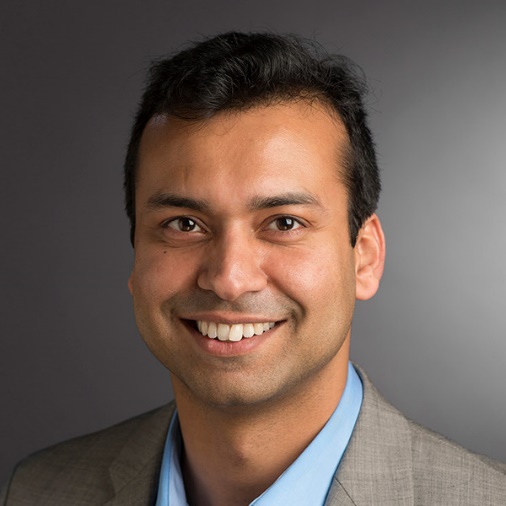}}]{Amlan Ganguly} (SM'21) is Professor and Department Head of Computer Engineering at Rochester Institute of Technology. Dr. Ganguly received his Bachelor of Technology from Indian Institute of Technology, Kharagpur and MS and PhD from Washington State University. His research interests are in interconnection networks for multi/many-core chips, Processing-in-Memory, Computing with DNA molecules, Data Centers, Edge Computing, their security and sustainability. He is a senior member of IEEE.
\end{IEEEbiography}

\begin{IEEEbiography}[{\includegraphics[width=1in,height=1.25in,clip,keepaspectratio]{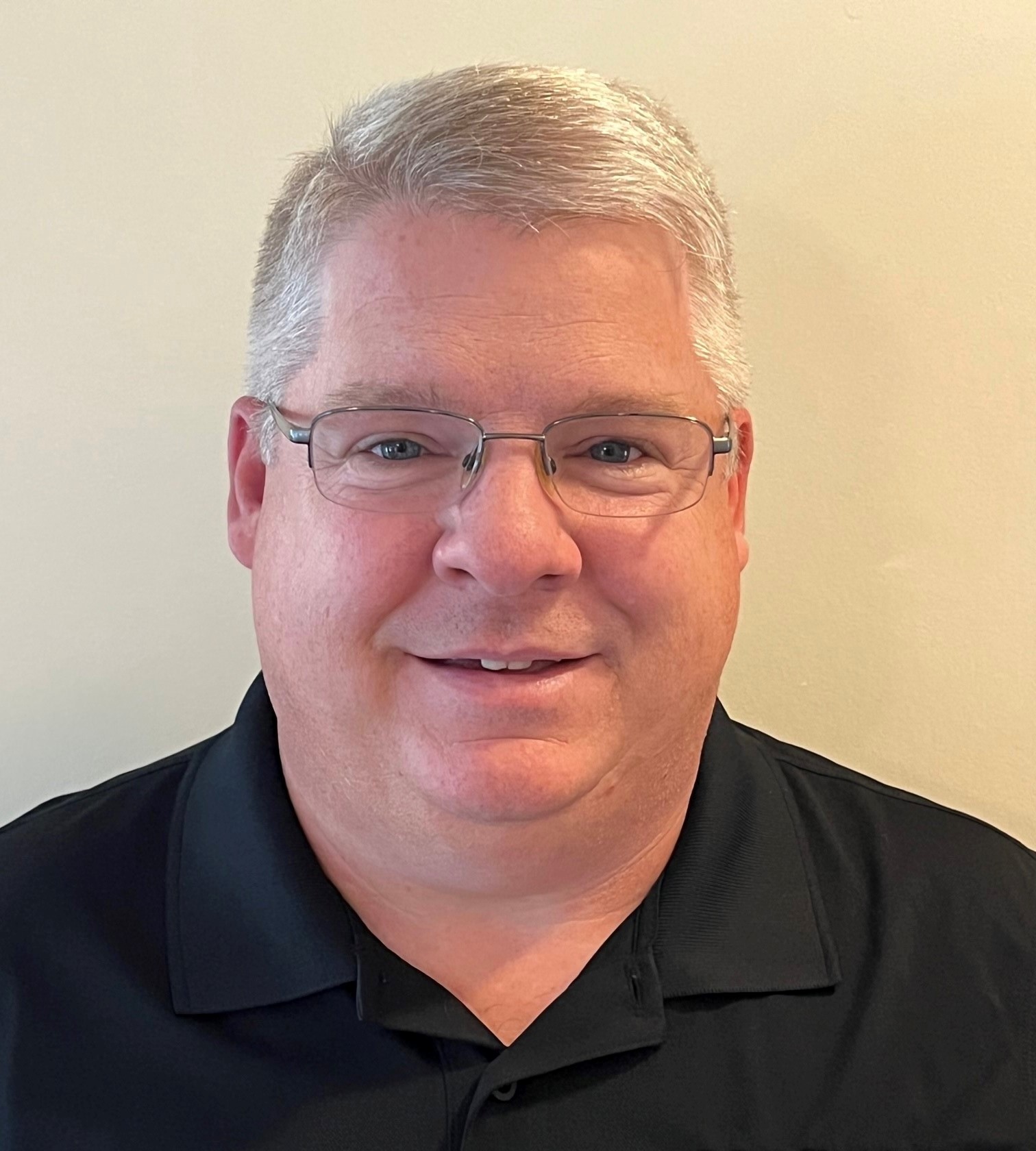}}]{Michael E. Kuhl,} is a Professor in the Department of Industrial and Systems Engineering at Rochester Institute of Technology in Rochester, New York. He received his PhD degree in industrial engineering from North Carolina State University in 1997. His research interests include simulation modeling and analysis, digital twins, Industry 4.0/5.0, human-robot interaction, and the design and development of smart material handling systems. 
\end{IEEEbiography}

\begin{IEEEbiography}[{\includegraphics[width=1in,height=1.25in,clip,keepaspectratio]{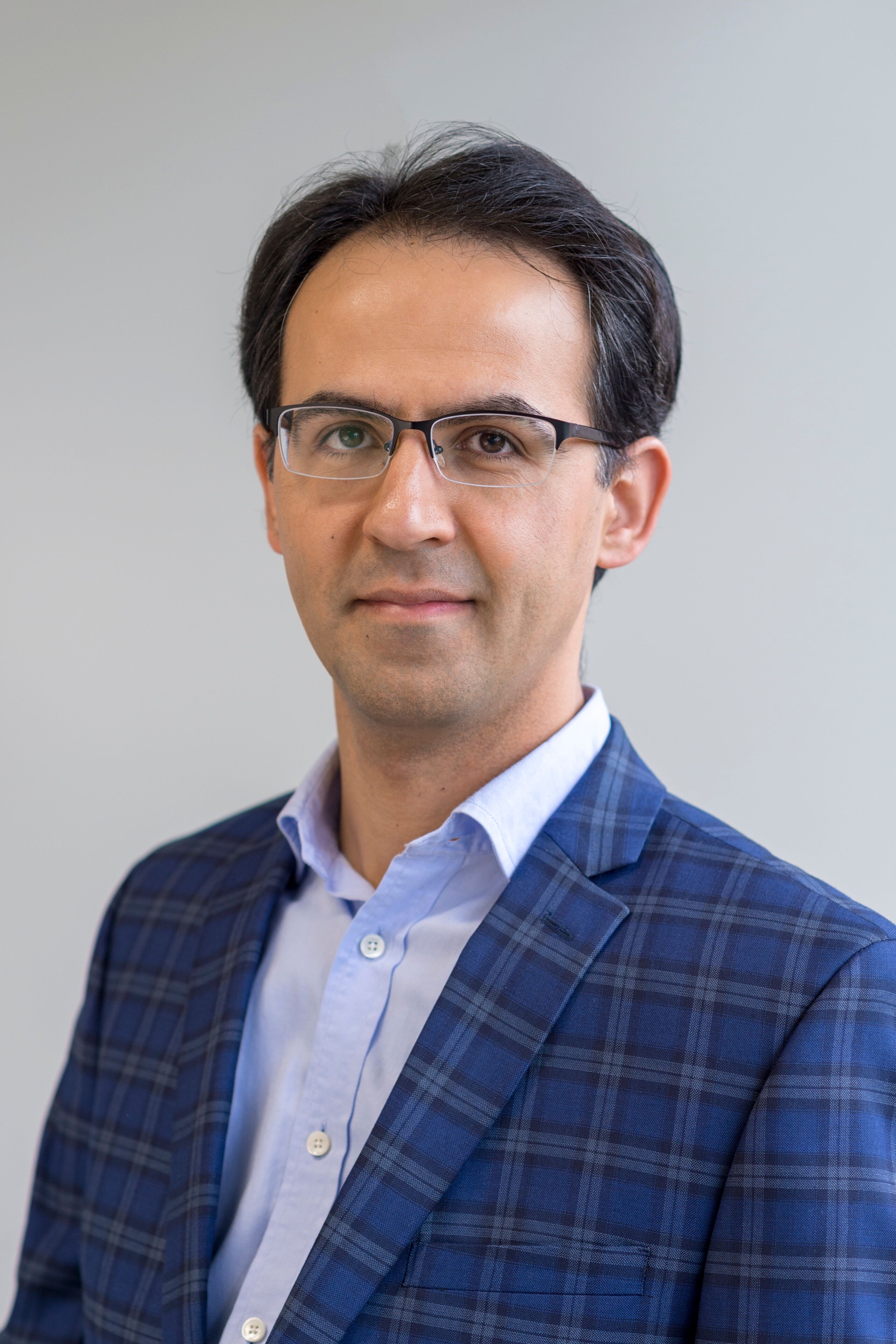}}]{Ehsan Rashedi} is an Assistant Professor in the Department of Industrial and Systems Engineering at Rochester Institute of Technology. He directs the Biomechanics and Ergonomics Lab, where his research focuses on human-centered system design, occupational biomechanics, and human–robot interaction. His work integrates wearable sensing, computer vision, and AI to assess human motion, improve workplace safety, and enhance collaboration between humans and autonomous systems.
\end{IEEEbiography}

\begin{IEEEbiography}[{\includegraphics[width=1in,height=1.25in,clip,keepaspectratio]{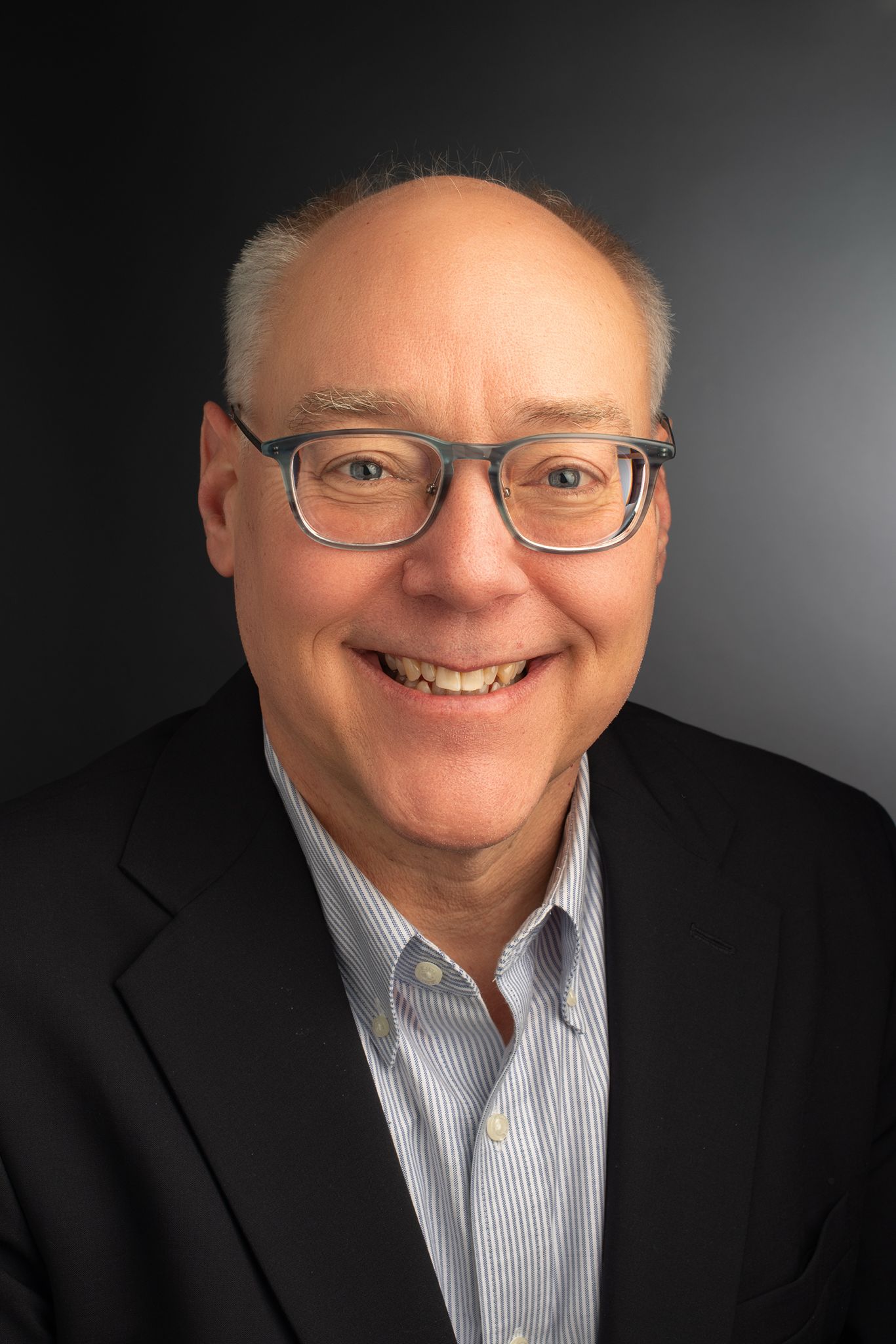}}]{Clark Hochgraf} is Director of Artificial intelligence for Rebuild Manufacturing. He previously served as generative AI faculty fellow, Miller chair for international education, and associate professor at Rochester Institute of technology where he researched innovative methods for human to autonomous mobile robot communication using gesture languages, wireless localization methods, and hands-on experiential learning for control systems and digital signal processing courses. He has a PhD in electrical engineering from the University of Wisconsin Madison, and a bachelors in electrical engineering from SUNY Buffalo.
\end{IEEEbiography}

\end{document}